\documentclass[twocolumn,aps,prd,amsmath,amssymb,nofootinbib]{revtex4-1}
\pdfoutput=1

\usepackage{graphicx,bm}

\newcommand{\beq}{\begin{equation}}
\newcommand{\bea}{\begin{eqnarray}}
\newcommand{\eeq}{\end{equation}}
\newcommand{\eea}{\end{eqnarray}}

\newcommand{\N}{N_{\mbox{\tiny{tot}}}}
\newcommand{\E}{P_{\mbox{\tiny{tot}}}}
\newcommand{\I}{\mathrm{i}}

\begin{document}

\title{Quantum and Classical Behavior in Interacting Bosonic Systems}

\author{Mark P.~Hertzberg}
\affiliation{Institute of Cosmology \& Dept.~of Physics and Astronomy,\\ Tufts University, Medford, MA 02155, USA}

\date{\today}

\begin{abstract}
It is understood that in free bosonic theories, the classical field theory accurately describes the full quantum theory when the occupancy numbers of systems are very large. However, the situation is less understood in interacting theories, especially on time scales longer than the dynamical relaxation time. Recently there have been claims that the quantum theory deviates spectacularly from the classical theory on this time scale, even if the occupancy numbers are extremely large. Furthermore, it is claimed that the quantum theory quickly thermalizes while the classical theory does not. The evidence for these claims comes from noticing a spectacular difference in the time evolution of expectation values of quantum operators compared to the classical micro-state evolution. If true, this would have dramatic consequences for many important phenomena, including laboratory studies of interacting BECs, dark matter axions, preheating after inflation, etc. In this work we critically examine these claims. We show that in fact the classical theory can describe the quantum behavior in the high occupancy regime, even when interactions are large. The connection is that the expectation values of quantum operators in a single quantum micro-state are approximated by a corresponding classical {\em ensemble average} over many classical micro-states. Furthermore, by the ergodic theorem, a classical ensemble average of local fields with statistical translation invariance is the spatial average of a single micro-state. So the correlation functions of the quantum and classical field theories of a single micro-state approximately agree at high occupancy, even in interacting systems. Furthermore, both quantum and classical field theories can thermalize, when appropriate coarse graining is introduced, with the classical case requiring a cutoff on low occupancy UV modes. We discuss applications of our results.
\end{abstract}
\let\thefootnote\relax\footnotetext{Electronic address: {\tt mark.hertzberg@tufts.edu}}

\maketitle

\section{Introduction}

Interacting bosons play an important role in many regimes, such as condensed matter systems of $^4$He, superfluidity, to particle physics of Higgs, W-bosons, and to various cosmological applications, such as inflation, preheating, etc. One especially interesting application is to dark matter axions \cite{Preskill:1982cy,Abbott:1982af,Dine:1982ah}. Since axions are very light (a  QCD axion might have mass $m_a\sim 10^{-5}$\,eV or so) their density and hence occupancy numbers are huge, so their bosonic character is important.

In general the behavior of interacting quantum particles is extraordinarily difficult to calculate, except in special circumstances. One important regime is the ultra-quantum regime of bosons at very high occupancy. In this regime, one would certainly not use classical particle physics. Instead one appears to require the full treatment of the many particle Schr\"odinger equation. However, it is well understood, at least for free or nearly free systems, that in such a regime, a different type of classical approximation emerges: namely the classical field theory. Such a treatment is often employed for interacting systems, for example to understand the basic dynamics of Bose-Einstein condensates \cite{BECbook}, to run lattice simulations of preheating \cite{Felder:2006cc,Amin:2014eta}, and to compute the evolution of the whole universe governed by scalar dark matter \cite{Marsh:2015xka}.

One might be concerned that this treatment is a little too naive. It is possible that the classical field theory is no longer applicable on some time scale associated with interactions. Indeed there is inevitably at least one very noticeable difference between interacting quantum and classical systems: In the classical system, the state evolves uniquely. While in the quantum system, even states with rather well defined initial values for, say, the field and its conjugate momentum, will inevitably have these quantities become less and less well defined as wavefunctions spread. This spread is most dramatic in interacting theories, especially among systems that exhibit chaos. So even for bosons at high occupancy, several modes can interact with one another causing a huge spread in the wavefunction and calling into question any possible role of the classical approximation.

Although the above argument was not quite the motivation, related concerns appeared in the recent work of Ref.~\cite{Sikivie:2016enz} (earlier discussion appears in Refs.~\cite{Erken:2011dz,Sikivie:2009qn}), where it was claimed that quantum and classical interacting theories deviate on a dynamical time scale $\tau$, even at high occupancy. At first sight this claim seems implausible, since the time scale $\tau$ is a property that can be defined purely within the classical theory. So it is very strange that classical physics should fail on a time scale independent of Planck's constant $\hbar$. In fact there exists an interesting literature on this subject, including the work of Refs.~\cite{Bodeker:1996wb,Bodeker:1997yy} where it is shown that agreement between quantum and classical thermal systems is in fact correct to order $\sim\hbar^2$ in anharmonic systems. Other notable work includes Ref.~\cite{Mueller:2002gd} where it is shown that the Boltzmann equation and classical field theory are related at high occupancy. Thermalization in classical field theory has been studied in Ref.~\cite{Boyanovsky:2003tc} and work on simulating quantum fields using classical physics includes \cite{Hirayama:2005ba}.

Nevertheless numerical studies in Ref.~\cite{Sikivie:2016enz} appear to indeed justify the claim that classical physics generically fails in interacting systems. The evidence presented was to consider a system that begins in a state of definite particle number $|\{N_i\}\rangle$ for a set of modes labelled $i$. This state was evolved by the Heisenberg equation of motion for a specific toy model for a choice of $N_i$. On the other hand, a related classical problem was also studied, where the annihilation operators $\hat{a}_i$ were replaced by $\mathbb{C}$ numbers $\psi_i$ (as is usual for classical field theory), whose initial magnitudes were set to $\sqrt{N_i}$ and phases set to $\theta_i=0$. The classical values of $N_i(t)$ were shown to deviate spectacularly from the quantum expectation values of $\langle \hat{N}_i(t)\rangle$ on time scales longer than $\sim\tau$. Furthermore it was shown that the classical values of $N_i(t)$ kept oscillating significantly throughout the simulation, without settling down (see Figure \ref{TimeEvolution} top left panel), while the quantum expectation values $\langle \hat{N}_i(t)\rangle$ settled down to near constant values at late times. It was thus suggested that the quantum system has relaxed to thermal equilibrium, while the classical system has not.

In this paper, we take a critical view of these conclusions. We point out that the appropriate comparison between quantum and classical is not to compare a quantum expectation value in a quantum state to the evolution of one very special classical micro-state (one with $\theta_i=0$ initially), but to an {\em ensemble} of classical states. Indeed a classical micro-state, which of course oscillates forever in a closed frictionless system, should not be compared to a quantum coarse grained quantity, namely an expectation value. In fact both classical and quantum micro-states oscillate wildly forever, while averaged values in both theories can settle down.

We provide numerical and analytical evidence that the quantum expectation values of the occupancy numbers are approximately given by the classical ensemble average of classical micro-states with initial phases drawn randomly from a uniform distribution (for a sample, see Figure \ref{TimeEvolution} top right panel). This is appropriate as the initial quantum states are chosen to be states of definite occupancy number and hence they have completely unspecified phases. This is in contrast to the work of Ref.~\cite{Sikivie:2016enz} which focusses only on the special $\theta_i=0$ case, which is not connected to the quantum state in any meaningful way.

We show that expectation values of classical states also settle down and approach the equilibrium values (see Figure \ref{TimeEvolution} bottom right panel) in the same way the quantum expectation values do. Hence both classical and quantum treatments can exhibit thermalization if the thermodynamic limit is taken and appropriate coarse graining is introduced. 

Finally we comment on an application to continuum field theory. In this case the classical ensemble average can be replaced by a spatial average of a single classical micro-state by the ergodic theorem (assuming an underlying translationally invariant distribution). Hence in this way, even a single classical micro-state can approximate correlation functions of the quantum theory, despite the quantum spreading of the wavefunction. This ``spreading" is captured by the ensemble or ergodicity. As an application, this means that the correlation length of dark matter axions is captured, at least approximately, by the classical theory.
(Of course in the continuum field theory, one should be concerned about the UV behavior of the classical theory involving low occupancy modes, which require artificial regulation. But this is not relevant to the work of Ref.~\cite{Sikivie:2016enz} which focusses only on a handful of finite frequency modes, all at high occupancy.)

Our paper is organized as follows: 
In Section \ref{Models} we introduce a class of interacting models of bosons. 
In Section \ref{Classical} we provide numerical results from evolving and ensemble averaging the classical evolution.
In Section \ref{Thermal} we compute the thermal averages and compare to numerics.
In Section \ref{Quantum} we show analytically that the expectation values match.
In Section \ref{Correlation} we discuss the implications for correlation functions and correlation lengths in local field theories.
Finally, in Section \ref{Discussion} we present a discussion.

\section{Bosonic Models}\label{Models}

Our primary motivation comes from systems of $\N$ bosons with a conserved particle number. This usually emerges in the non-relativistic limit (important examples include laboratory studies of $^4$He due to conservation of baryon number and dark matter axions due to small annihilation cross sections). Furthermore, we will focus on $2\to2$ scattering processes as these tend to dominate in the non-relativistic theory. A set of relevant interactions are of the standard form
\beq
\hat{H}=\sum_{i=1}^{\N}{\hat{p}_i^2\over2 m}+\sum_{i<j}^{\N}V(\hat{\bf x}_i-\hat{\bf x}_j).
\eeq
Some important examples include gravitation with $V=-Gm^2/|\hat{\bf x}_i-\hat{\bf x}_j|$ and $\lambda\,\phi^4$ theory with $V\sim\lambda\,\delta(\hat{\bf x}_i-\hat{\bf x}_j)/m^2$.

Since we are interested in bosons at high occupancy, it is useful to pass to the second quantized language using creation and annihilation operators ${\hat a}_i^\dagger,\, {\hat a}_i$ (where index $i$ labels each mode with wave-vector ${\bf k}_i$). For a discrete set of momenta the above Hamiltonian can be re-written as
\beq
\hat{H}=\sum_i \omega_i\, \hat{a}^\dagger_i\hat{a}_i+{1\over4}\sum_{ijkl}\Lambda_{ij}^{kl}\,\hat{a}^\dagger_i\hat{a}^\dagger_{j}\hat{a}_{k}\hat{a}_{l},
\label{Hamiltonian}\eeq
where $\omega_i=|{\bf k}_i|^2/(2m)$ is the frequency of each mode. The potential is re-organized into some collection of coefficients encoded in the couplings $\Lambda_{ij}^{kl}$ (related to the Fourier transform of $V$). The conservation of momentum requires $\Lambda_{ij}^{kl}$ to only be non-zero when the momenta associated with each index satisfy ${\bf k}_i+{\bf k}_j={\bf k}_k+{\bf k}_l$. The hermiticity of the Hamiltonian requires $\Lambda_{lk}^{ji*}=\Lambda_{ij}^{kl}$. Note that the convention of eq.~(\ref{Hamiltonian}) is to allow over-counting of indices, so we can take without loss of generality $\Lambda_{ij}^{kl}=\Lambda_{ji}^{kl}=\Lambda_{ij}^{lk}=\Lambda_{ji}^{lk}$.

In principle we would like to study a huge set of wave-vectors to properly address the continuum theory. However, the numerical evolution is very difficult. It suffices for the present purposes to simply study a toy problem built out of only a handful of oscillators. We will follow the interesting and useful model of Ref.~\cite{Sikivie:2016enz} (earlier introduced in Ref.~\cite{Erken:2011dz}). 

The details of the model are as follows: Only 5 modes are included, labelled by $i=1,\ldots,5$, with $k_i\propto i$ just a scalar. The non-relativistic dispersion relation is replaced by a linear dispersion relation $\omega= k$. So the frequencies of the 5 oscillators are taken to be integer multiples of a fundamental frequency $\omega_0$ as 
\beq
\omega_i=i\,\omega_0,\,\,\,i=1,\ldots,5.
\label{frequencies}\eeq
 The only non-zero independent couplings are taken to all equal some overall strength of interaction $\Lambda_0$ as $\Lambda_{14}^{23}=\Lambda_{15}^{24}=\Lambda_{25}^{34}=\Lambda_{22}^{13}/2=\Lambda_{33}^{24}/2=\Lambda_{33}^{15}/2=\Lambda_{44}^{35}/2=\Lambda_0$. As a concrete example of initial conditions, Ref.~\cite{Sikivie:2016enz} took the initial quantum state to have definite occupancy numbers of $|\{N_i\}\rangle=|12,25,4,12,1\rangle$. The coupling is chosen to be $\Lambda_0=\omega_0/10$, while the inverse frequency $\omega_0^{-1}$ can be used a unit of time. 

In Ref.~\cite{Sikivie:2016enz} the system was evolved under the Heisenberg equations of motion following from the Hamiltonian in (\ref{Hamiltonian}). They then outputted the expectation value of the occupancy operators $\langle\hat{N}_i(t)\rangle=\langle\hat{a}_i^\dagger(t)\,\hat{a}_i(t)\rangle$ over time, finding that on a fairly short time scale the occupancy number expectation values settled towards their equilibrium values.

\section{Classical Behavior}\label{Classical}

Let us now study this problem carefully within the framework of classical physics. The classical theory arises from replacing the annihilation operators $\hat{a}_i$ by $\mathbb{C}$ numbers $\psi_i$. The corresponding classical equation of motion comes from Hamilton's equations using the fact that $\I\,\psi^*$ is the momentum conjugate to $\psi$. This gives the set of ODEs
\beq
\I\,\dot{\psi}_i = \omega_i\,\psi_i+{1\over2}\sum_{jkl}\Lambda_{ij}^{kl}\,\psi_j^*\psi_k\psi_l.
\label{CEOM}\eeq
The complex variables $\psi_i$ are specified by a magnitude and a phase that we can write as 
\beq
\psi_i(t)=\sqrt{N_i(t)}\,e^{\I\theta_i(t)},
\eeq 
where $N_i(t)$ has an interpretation as an ``occupancy number" in the quantum theory. 

According to Ref.~\cite{Sikivie:2016enz} the classical analogue of the initial quantum state of definite occupancy numbers $|\Psi(t=0)\rangle=|\{N_i\}\rangle$ is to choose $\psi_i(t=0)=\sqrt{N_i}$ with all phases vanishing $\theta_i(t=0)=0$. However, this is not the right analogue. Since the quantum state has an {\em unspecified} phase, then we should not try to connect it to a classical state of a specific value of $\theta_i=0$. Instead the classical analogue is an {\em ensemble} of states with starting values of $\theta_i$ drawn independently and randomly from the uniform distribution on the domain
\beq
\theta_i(t=0)\in[0,2\pi),\,\,\,\,i=1,\ldots,5.
\eeq
In Section \ref{Quantum} we will explain why this is the appropriate ensemble of initial states. These initial states can then each be evolved under the classical equations of motion (\ref{CEOM}). Finally we can output the expectation value of the modulus square of the field after ensemble averaging over $s$ initial sets of phases, which we denote  $\langle N_i\rangle_{ens}^{(s)}$. 

In the classical problem, we can scale out the overall number of particles $\N$, which is only important in the quantum problem. We can define the fractional occupancy numbers $n_i(t)\equiv N_i(t)/\N$, which satisfy the conservation law 
\beq
\sum_i n_i(t)=1,
\eeq
and we can define a dimensionless coupling parameter 
\beq
\tilde\Lambda\equiv{\Lambda_0\,\N\over\omega_0}.
\eeq
Following Ref.~\cite{Sikivie:2016enz} we choose the following set of initial values for $\{n_i\}$: $\{12/54,25/54,4/54,12/54,1/54\}$ and coupling: $\tilde\Lambda=54/10$. There is a dynamical time scale $\tau$ in the problem, which is roughly $\tau^{-1}\sim\Lambda_0\,N_i\sqrt{h}\sim 3\,\omega_0$, where $h=7$ is the number of interaction terms in the Hamiltonian.

\begin{center}
\begin{figure}[h!]
\vspace{2cm}
\includegraphics[width=8.4cm]{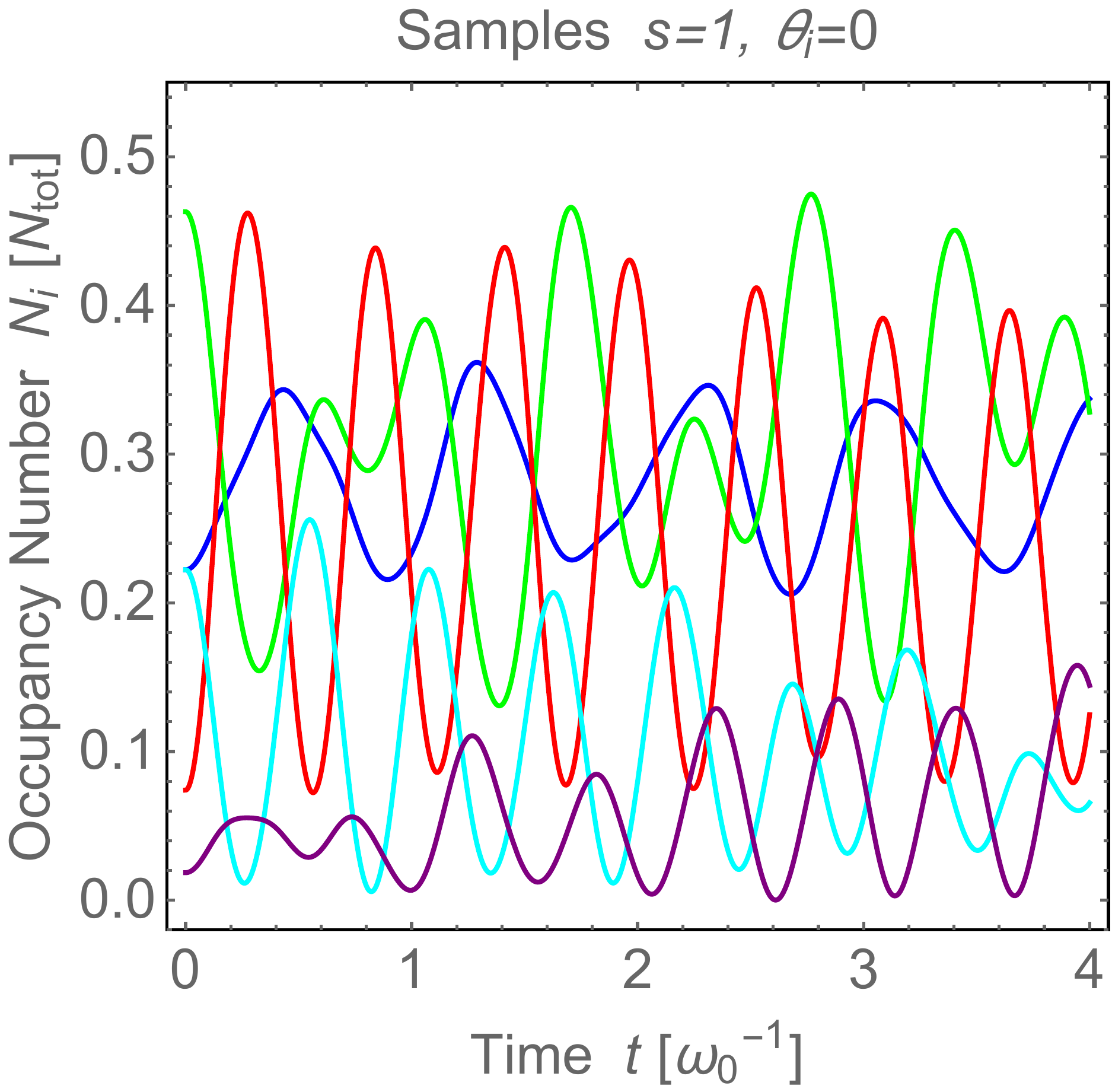}\,\,\,\,\,\,\,\,\,\,
\includegraphics[width=8.4cm]{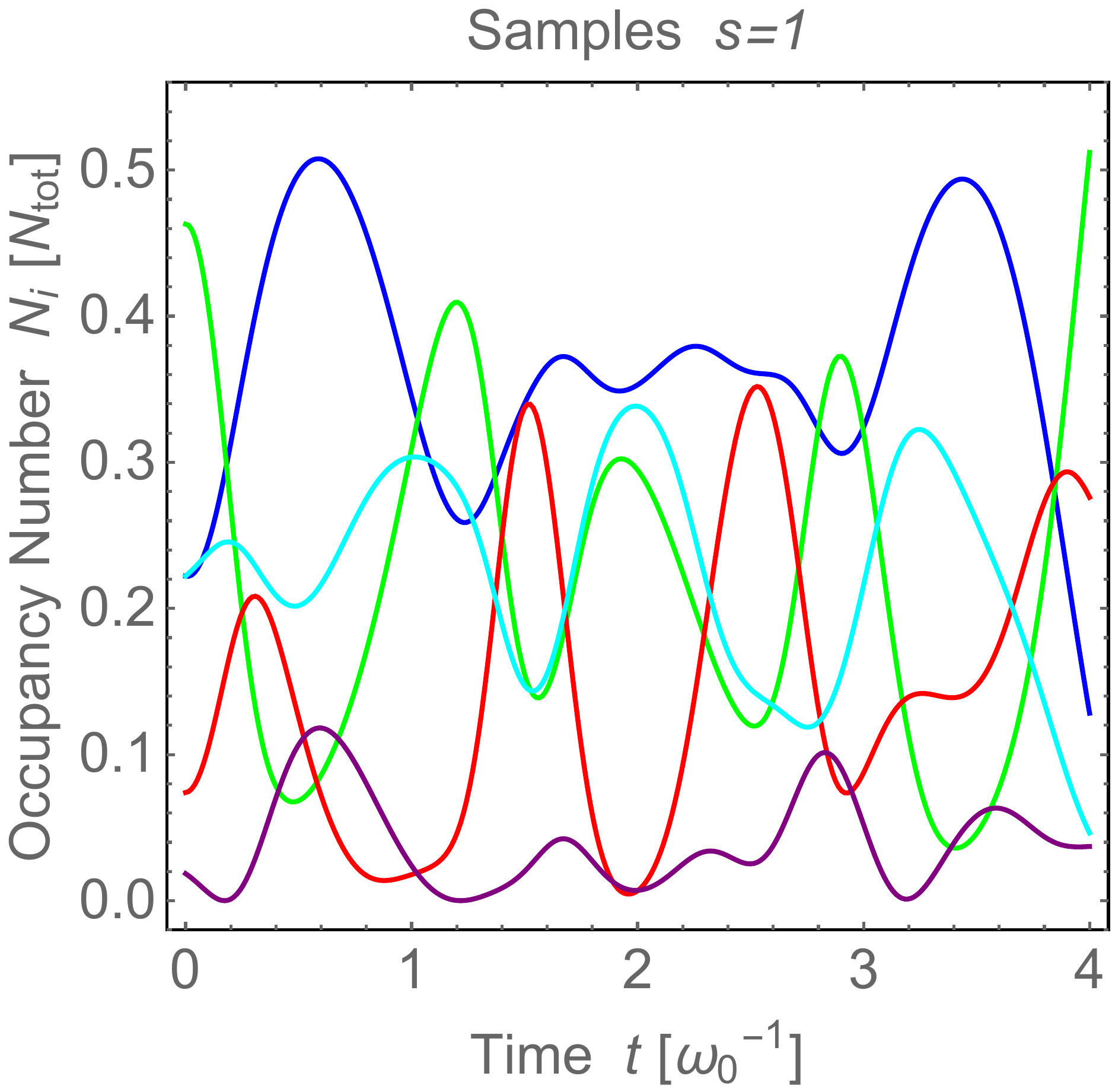}\,\,\,\,\,\,\\
\vspace{0.4cm}
\includegraphics[width=8.4cm]{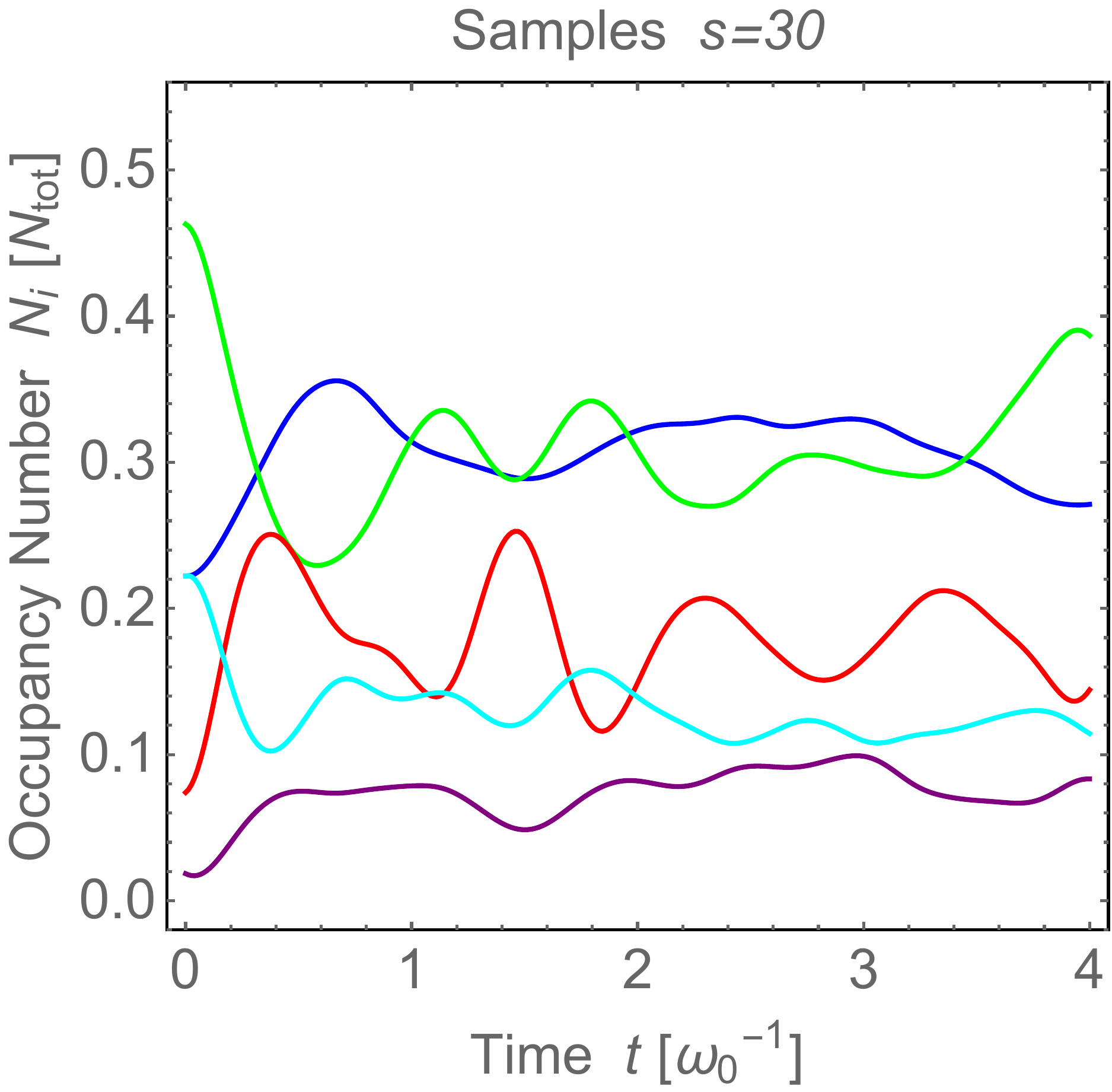}\,\,\,\,\,\,\,\,\,\,
\includegraphics[width=8.4cm]{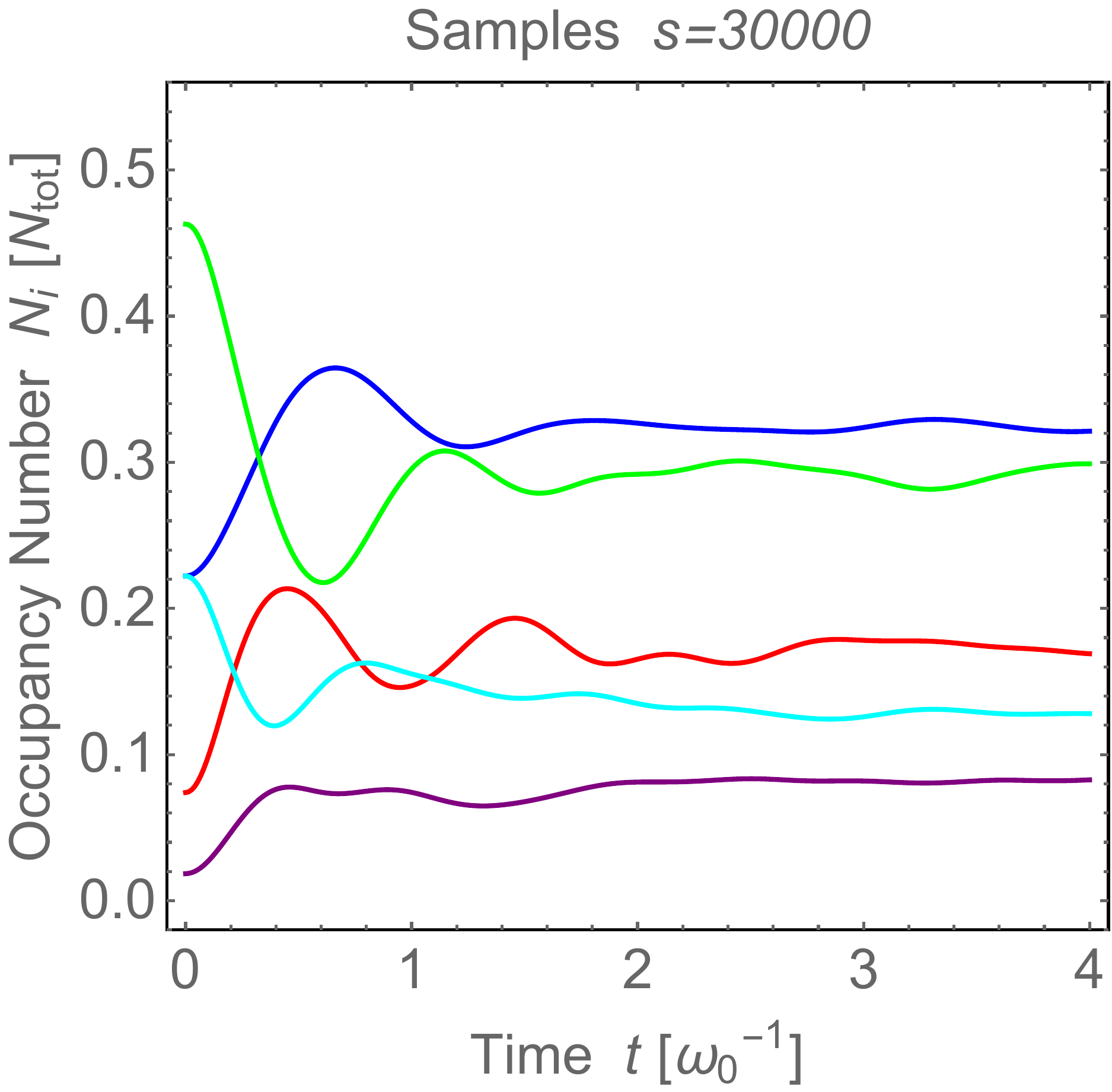}\!\!
\includegraphics[width=0.3156cm]{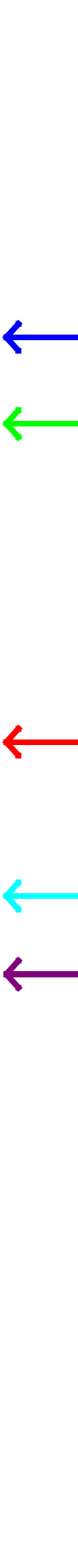}
\caption{Evolution of the classical occupancy numbers $\langle N_i\rangle_{ens}^{(s)}$ (in units $\N$) over time $t$ (in units $\omega_0^{-1}$) after averaging $s$ ensembles with phases $\theta_i$ to set the initial conditions. Top left panel $s=1$ (special choice of phases $\theta_i=0$); top right panel $s=1$ (random phases); bottom left panel $s=30$ (random phases); bottom right panel $s=30000$ (random phases) with thermal equilibrium values denoted by arrows. The curves correspond to each of the 5 oscillators: $\omega_1=\omega_0$ is blue; $\omega_2=2\,\omega_0$ is green; $\omega_3=3\,\omega_0$ is red; $\omega_4=4\,\omega_0$ is cyan; $\omega_5=5\,\omega_0$ is purple. The initial values of $\{N_i/\N\}$ are set to $\{12/54,25/54,4/54,12/54,1/54\}$ and the coupling is $\Lambda_0\,\N/\omega_0=54/10$. At high occupancy, the expectation values of the quantum occupancy number operators $\langle\hat{N}_i\rangle$ are approximated by the true classical ensemble averages $s\to\infty$. So $\langle\hat{N}_i\rangle$ at high occupancy is approximated by the bottom right panel $s=30000$.} 
\label{TimeEvolution}\end{figure}
\end{center}

We have solved this system of classical equations numerically with results presented in Figure \ref{TimeEvolution}. In the top left panel we output the special case in which all the phases are set $\theta_i(t=0)=0$. This state is highly non-generic, but was used as representing the classical theory in Ref.~\cite{Sikivie:2016enz}. In the top right panel we output a much more generic case in which the $\theta_i(t=0)$ are chosen randomly. This evolution exhibits considerably more chaos than the special case. In the bottom panels we then pick $s=30$ and $s=30000$ random sets of initial $\theta_i$ and average the solutions. Even for $s=30$ in lower left panel we see somewhat less variation compared to the top right panel. For $s=30000$ we are essentially in the limit in which we have achieved the true ensemble average 
\beq
\langle N_i\rangle_{ens}=\langle N_i\rangle_{ens}^{(\infty)}\approx\langle N_i\rangle_{ens}^{(30000)}.
\eeq
\onecolumngrid
\twocolumngrid

The ensemble averaged evolution is seen to be similar to the evolution of the quantum expectation value $\langle\hat{N}_i\rangle$ of the quantum state that was computed numerically in Ref.~\cite{Sikivie:2016enz} (we refer the reader to those figures for comparison). Although the quantum case can only be computed efficiently for finite $N_i$, we will show in Section \ref{Quantum} that in the high $N_i$ regime, the two answers will approximately agree
\beq
\langle \hat{N}_i(t)\rangle\approx\langle N_i(t)\rangle_{ens}.
\eeq
Furthermore, at a fixed time $t$, we expect these to converge in a limit in which we take $\N\to\infty$ and $\Lambda_0\to0$, while keeping $\tilde\Lambda$ finite.

\section{Thermal Averages}\label{Thermal}

The ensemble averaged classical occupancy numbers roughly approach some equilibrium value (presumably one needs to include a large number of oscillators to truly be in the thermodynamic limit and reach true equilibrium). One can enquire whether they approximate the thermal equilibrium values $\bar{N}_i$. Computing the exact thermal equilibrium is ordinarily difficult as we would need to perform statistical mechanics of a nonlinear system. However, in this special toy model, we can focus on the free theory Hamiltonian $\hat{H}_0$ which is the conserved momentum $\E$ in a theory with dispersion $\omega_i=k_i$. (In other cases, one sometimes just approximates equilibrium by using the free theory. This can fail, such as for attractive interactions in the continuum field theory \cite{Guth:2014hsa}).

In Ref.~\cite{Sikivie:2016enz} it was claimed that thermal equilibrium in the classical theory would mean equipartition of (free theory) energy into each of the 5 oscillators: $\bar{E}_{0i}=\bar{N}_1\,\omega_1=\bar{N}_2\,\omega_2=\bar{N}_3\,\omega_3=\bar{N}_4\,\omega_4=\bar{N}_5\,\omega_5$. Using eq.~(\ref{frequencies}) this immediately gives a set of values for $\bar{N}_i$ which disagree considerably with the late time values seen in Figure \ref{TimeEvolution} lower right panel. However it is incorrect to use equipartition of energy in this case because it ignores the fact that the number of particles is fixed.

Instead the correct treatment of classical thermal equilibrium in this framework is to use the micro-canonical ensemble with momentum and number of particles fixed. So the macro-state is specified by $M=\{\E,\N\}$. The entire set of allowed micro-states are any $\mu=\{N_i,\theta_i\}$ that satisfy the 2 constraints
\beq
\E=\sum_i N_i\,k_i\,\,\,\,\,\, \mbox{and}\,\,\,\,\, \N=\sum_i N_i.
\eeq
Thermal averages for the occupancy numbers are then given by the following integrals
\beq
\bar{N}_i ={\int\!\Big{[}\prod_j dN_j\Big{]} N_i\,\delta(\E-\sum_j N_j\,k_j)\,\delta(\N-\sum_j N_j)\over
\int\!\Big{[}\prod_j dN_j\Big{]}\delta(\E-\sum_j N_j\,k_j)\,\delta(\N-\sum_j N_j)}
\label{Mean}\eeq
and similarly for other moments such as $\bar{N_i^2}$.

For the example studied in Figure \ref{TimeEvolution}, the input momentum to particle number ratio is $\E/\N=127\omega_0/54$. By carrying out these integrals we obtain the following set of thermal equilibrium values in the classical theory:
\bea
\{\bar{n}_i\}&\approx&\{0.325,\,0.293,\,0.175,\,0.118,\,0.089\},  \\
\{\sigma_{n_i}\}&\approx&\{0.145,\,0.211,\,0.132,\,0.088,\,0.066\},
\eea
where $\bar{n}_i\equiv\bar{N}_i/N$ is the mean fractional occupancy in each mode and $\sigma_{n_i}^2\equiv \bar{n_i^2}-\bar{n}_i^2$ is the variance. The thermal averages $\{\bar{n}_i\}$ are indicated by arrows in the lower right panel of Figure \ref{TimeEvolution}. We see that they match the late time ensemble averages of the simulation quite well. We have also checked that they match a long time temporal average of a single micro-state quite well too. Also note that the size of the $\sigma_{n_i}$ is of the same order as $\bar{n}_i$, so the fluctuations are large. 

In the quantum theory, the thermal averages come from an almost identical calculation to eq.~(\ref{Mean}), but with integrals replaced by discrete sums. Hence it is obvious that these two approaches agree at high occupancy where the discrete sum may be approximated by an integral. So the quantum fluctuations are equally large. This is due to the spreading of the (occupancy basis) wavefunction and is captured by the spread in members of the classical ensemble.

\section{Quantum to Classical Connection}\label{Quantum}

Here we would like to explain why the classical ensemble average reproduces the quantum expectation values. To begin, consider the classical equations of motion (\ref{CEOM}). We denote the initial values as $\psi_i(t=0)=\phi_{i}$ and we can solve this system of equations as a Taylor series in time. If we form the modulus square of $\psi_i$, this provides the Taylor series of the classical occupancy $N_i(t)$, which takes the form
\beq
N_i(t) = \phi_i^*\phi_i-{t\over2}\Big{[}\I\sum_{jkl}\Lambda_{ij}^{kl}\,\phi_i^*\phi_j^*\phi_k\phi_l+c.c\Big{]} +\ldots,
\label{Nsoln}\eeq
where the coefficient of $t^p$ in the Taylor expansion is a polynomial in $\phi_i,\,\phi_j^*$ of order $2p+2$.

On the other hand we can also compute the time evolution in the quantum theory. The Heisenberg equation of motion is similar to eq.~(\ref{CEOM}), but with the replacement $\psi_i\to\hat{a}_i$ and $\psi_i^*\to\hat{a}^\dagger_i$. Let us denote the initial values for these operators as $\hat{a}_i(t=0)=\hat{b}_i$. In principle we can solve this system of equations, but it is much more difficult due to the fact that the creation and annihilation operators do not commute
\beq
\hat{a}_i\,\hat{a}_j^\dagger=\hat{a}_j^\dagger\,\hat{a}_i+\delta_{ij}.
\label{commutation}\eeq
However there is tremendous simplification in the high occupancy regime. In this case the typical values of these operators are large, in the sense that expectation values $\langle\hat{a}_i^\dagger\hat{a}_i\rangle$ are large. So in this regime we do not need to be concerned about the $\delta_{ij}$ correction of (\ref{commutation}). This is a relative error of $\mathcal{O}(1/N_i)$. Hence we can freely commute these operators, which means that the structure of the solution reduces to precisely eq.~(\ref{Nsoln}), at each order, with $N_i(t)\to\hat{N}_i(t)$, $\phi_i\to\hat{b}_i$, and $\phi_i^*\to\hat{b}_i^\dagger$, plus $\mathcal{O}(1/N_i)$ relative corrections
\beq
\hat{N}_i(t) \approx \hat{b}^\dagger_i\hat{b}_i-{t\over2}\Big{[}\I\sum_{jkl}\Lambda_{ij}^{kl}\,\hat{b}^\dagger_i\hat{b}^\dagger_j\hat{b}_k\hat{b}_l+h.c\Big{]} +\ldots,
\label{NQsoln}\eeq
where, as above, the coefficient of $t^p$ in the Taylor expansion is a polynomial in $\hat{b}_i,\,\hat{b}_j^\dagger$ of order $2p+2$, with coefficients matching the classical case (\ref{Nsoln}).

Now we would like to compute the expectation value of $\hat{N}_i(t)$ in an initial state of definite occupancy $|\Psi(t=0)\rangle=|\{N_i\}\rangle$. We can compute this expectation value term by term in the series (\ref{NQsoln}) by using the standard ways in which creation and annihilation operators act on states
\bea
&&\hat{b}_i\,|N_1,\ldots,N_i,\ldots\rangle=\sqrt{N_i}\,|N_1,\ldots,N_i-1,\ldots\rangle, \,\,\,\,\,\,\,\,\,\,\,\,\,\,\,\,\,\,\,\,\\
&&\hat{b}_i^\dagger\,|N_1,\ldots,N_i,\ldots\rangle=\sqrt{N_i+1}\,|N_1,\ldots,N_i+1,\ldots\rangle.
\eea
Lets illustrate by taking the expectation value of a representative term in the expansion (\ref{NQsoln}), namely a collection of 4 operators. The orthogonality of the states $|\{N_i\}\rangle$ leads to
\bea
&&\langle\{N_i\}|\,\hat{b}^\dagger_j\hat{b}^\dagger_k\hat{b}_l\hat{b}_m\,|\{N_i\}\rangle \nonumber \\
&&\approx\sqrt{N_j\,N_k\,N_l\,N_m}\,(\delta_{jl}\delta_{km}+\delta_{jm}\delta_{kl}-\delta_{jklm}).\,\,
\label{4prod}\eea
In the classical case, the analogue is to perform an ensemble average over phases of the initial $\psi_i(t=0)=\phi_i=\sqrt{N_i}\,e^{\I\theta_i(t=0)}$, so the corresponding classical term has expectation value
\bea
&&\langle\phi_j^*\phi_k^*\phi_l\phi_m\rangle_{ens} \nonumber \\
&&=\sqrt{N_j\,N_k\,N_l\,N_m}\,\int \Big{[}\prod_i {d\theta_i\over2\pi}\Big{]}\,e^{\I(\theta_l+\theta_m-\theta_j-\theta_k)}.\,
\eea
Carrying out the integral gives exactly eq.~(\ref{4prod}). (It turns out that this particular contribution vanishes when inserted into eq.~(\ref{Nsoln}) or (\ref{NQsoln}) for the toy model, but this illustrates the basic idea). This correspondence carries over to all the various higher order terms in the series (\ref{Nsoln}) and (\ref{NQsoln}) (many of which do {\em not} vanish). Hence the final results are indeed expected to agree in the large $N_i$ limit where we take $\N\to\infty$ and $\Lambda\to 0$ holding their product finite.

For finite $N_i$, it is not obvious what is the time scale for departure of these theories, as the errors may grow in the higher order terms in the Taylor expansion. But the classical numerics presented earlier, with the comparison to the quantum numerics in Ref.~\cite{Sikivie:2016enz}, indicates that they quantitatively agree for times much longer than $\tau$, and we believe they roughly agree for extremely long times as they exhibit very similar equilibria.

\section{Correlation Functions}\label{Correlation}

If we consider the continuum field theory, this connection allows us to express position space correlation functions of local operators in terms of the corresponding classical averages. For example, we may be interested in the two-point correlation function $\langle \{N_i\}|\hat\psi^\dagger({\bf x},t)\hat\psi({\bf y},t)|\{N_i\}\rangle$, which ordinarily encodes the correlation length as the scale over which the correlations fall off. At high occupancy we can express this as
\bea
\langle \{N_i\} |\hat\psi^\dagger({\bf x},t)\,\hat\psi({\bf y},t)|\{N_i\}\rangle \approx \langle\psi^*({\bf x},t)\,\psi({\bf y},t)\rangle_{ens}.\,\,\,\,\,\,\,\,
\label{correlation}\eea

Now in some sense the quantum and classical theories still differ in so far as the left hand side involves the expectation value of a single state, while the right hand side involves ensemble averaging many states. However, this is not necessarily so. The classical ensemble average may be replaced by a spatial average of a single classical micro-state $\psi_\mu$ by the ergodic theorem (assuming an underlying translationally invariant distribution, as provided by inflation, for example)
\bea
&&\langle\psi^*({\bf x},t)\,\psi({\bf y},t)\rangle_{ens} \nonumber\\
&&= {1\over V}\!\int_V d^3z\,\psi^*_\mu({\bf x}+{\bf z},t)\,\psi_\mu({\bf y}+{\bf z},t),
\label{ergodic}\eea
where the initial value of the classical field in ${\bf k}$-space is taken to be $\psi_\mu({\bf k}_i,t=0)=\sqrt{N_i}\,e^{\I\theta_i}$, where $\theta_i$ is randomly chosen.
Together eqs.~(\ref{correlation},\,\ref{ergodic}) provide an important result: 
{\em For a quantum micro-state, initially specified by $|\{N_i\}\rangle$, and a classical micro-state, initially specified by $\mu=\{N_i,\theta_i\}$, the quantum and classical correlation functions (and correlation lengths) approximately agree at high occupancy for long times.} Also, under certain circumstances, a temporal form of the ergodic theorem may be applicable too.

\section{Discussion}\label{Discussion}

In the case of axion dark matter, it was shown in Ref.~\cite{Guth:2014hsa} that the correlation length according to the classical theory is small because the interactions (gravity and self-interactions) are attractive rather than repulsive. The current analysis shows that the classical result for attractive interactions carries over directly to the quantum theory too. Hence axion dark matter does not lead to long range correlations. Instead it can (at least partially) thermalize and lead to the formation of Bose-Einstein condensate clumps, such as Bose stars. For details see Refs.~\cite{Guth:2014hsa,Khlebnikov:1999qy,Khlebnikov:1999pt,Kolb:1993hw,Kolb:1993zz,Hertzberg:2010yz,Eby:2015hyx} and for related discussions see Refs.~\cite{Berges:2014xea,Davidson:2013aba,RindlerDaller:2011kx,Li:2013nal,Noumi:2013zga}.

Our new results indicate that the classical description of bosonic fields can be entirely adequate, even though quantum wavefunctions do spread appreciably in interacting systems. This has application to not only axions, but to preheating simulations, etc. Also, this behavior can be seen in several other familiar contexts. For example, if one considers interacting billiard balls on a frictionless table, the wavefunctions spread, especially after each collision, so the expectation values of each ball's position $\langle\hat{x}_i(t)\rangle$ settle down at late times, while the classical micro-state $x_i(t)$ oscillates wildly. It is understood that this doesn't prevent classical physics from remaining a useful description of billiard balls.

Another familiar example is that of fluids governed by the Navier-Stokes equation. At the level of the effective field theory, one can, in principle, quantize the fluid's density and momentum density to formulate a quantum theory. Here there is interesting nonlinear behavior, such as turbulence, that is captured accurately by the classical field theory. Again one does not need to be concerned that the quantum wavefunction of the fluid has spread out on time scales longer than some dynamical time.

In all these cases, one can, in principle, perform an ensemble averaging over some space of initial conditions and use the classical evolution to mimic the quantum expectation values. Moreover, when an ergodic theorem applies, some spatial or temporal average can simply be performed to capture this. 

In practice, even this step is often unnecessary, however, since decoherence provides an effective collapsing of the wavefunction. So one can essentially utilize the classical theory with a single history, bearing in mind that one should not attempt to predict the future trajectory with detailed precision in chaotic systems, but only to represent the basic character of what an individual observer might see. Furthermore, for certain special states, such as a BEC, a single classical field configuration is usually accurate in describing its behavior, as the fluctuations around the condensate $\delta\hat\psi$ are often small \cite{BECbook}.

Finally, let us remark on a special class of initial quantum states, namely coherent states, which are often thought of as the most classical. In this case, we expect a single classical micro-state to match the quantum expectation value on a time scale that is parametrically $\sim\tau\,\ln \bar{N}$, as one expects $\sim\ln \bar{N}$ collisions for the small initial quantum uncertainty $\sim1/\sqrt{\bar{N}}$ to grow to be $\mathcal{O}(1)$ \cite{Albrecht:2012zp} as the system of interacting oscillators is chaotic. That there should be improved agreement between classical and quantum in the high occupancy limit is essentially guaranteed by the Ehrenfest theorem. Contradicting this well established theorem, Ref.~\cite{Sikivie:2016enz} claimed that the time scale for agreement is still only $\sim\tau$. We believe this is an artifact of running simulations with occupancy numbers that are too small to see the $\sim\ln \bar{N}$ enhancement. Indeed in order to study the coherent state, Ref.~\cite{Sikivie:2016enz} used mean initial occupancy numbers of $|0,12,16,0,0\rangle$, whose average value is $\bar{N}=5.6$. This is not a particularly high occupancy number and so the parametric enhancement of $\sim \ln \bar{N}$ is only an $\mathcal{O}(1)$ change to the $\sim\tau$ estimate. Instead one would need to study much higher occupancy numbers to clearly establish the logarithmic enhancement beyond the dynamical time scale $\tau$. 
In any case, as is the main point of this paper, we believe ensemble averaging (or ergodicity) is still essential to mimic the quantum thermalization for times $t\gg\tau\,\ln \bar{N}$. We know that at late times these simple systems thermalize and exhibit the same thermal distribution in the high occupancy regime. So we are then assured to have agreement for both early and intermediate times, due to Ehrenfest theorem, and at late times, due to similar thermalization after averaging. For coherent states, we can imagine some procedure of drawing the starting values of $N_i$ and $\theta_i$ from a distribution of relative widths $\mathcal{O}(1/\sqrt{N_i})$ around their starting mean values and then ensemble averaging.

\acknowledgments
We would like thank Lucas Kocia, Peter Love, Ali Masoumi, Mohammad Namjoo, Ken Olum, Alex Vilenkin, and especially Alan Guth for helpful discussions. We would like to thank the Tufts Institute of Cosmology for support.

\end{document}